\documentclass[sigconf]{acmart}

\usepackage{amssymb}
\usepackage{amsfonts}
\usepackage{algorithmic}
\usepackage{graphicx}
\usepackage{subfigure}
\usepackage{float}
\usepackage{wrapfig} %
\usepackage{xcolor} %
\usepackage{makecell} %
\usepackage{booktabs} %
\usepackage{amsmath}
\usepackage{accents} %
\usepackage{statex} %
\usepackage[normalem]{ulem} %
\usepackage{enumitem} %
\usepackage{multirow}
\usepackage[nomargin,inline,marginclue,draft]{fixme}
\usepackage{balance}
\usepackage{bm}
\usepackage{hyperref}
\usepackage[T1]{fontenc}
\usepackage{changepage} %
\usepackage{setspace} %
\usepackage{mathrsfs} %
\usepackage{verbatim} %
\usepackage{diagbox} %
\usepackage{pdftexcmds} %
\usepackage{catchfile} %
\usepackage{ifluatex} %
\usepackage{ifplatform} %
\usepackage{threeparttable} %
\usepackage{comment} %
\usepackage{aecompl} %

\usepackage{amsmath,amsfonts,bm}

\def\eqref#1{equation~\ref{#1}}

\def\1{\bm{1}}

\def\ve{{\bm{e}}}

\def\mE{{\bm{E}}}

\def\mM{{\bm{M}}}

\def\mR{{\bm{R}}}

\DeclareMathAlphabet{\mathsfit}{\encodingdefault}{\sfdefault}{m}{sl}
\SetMathAlphabet{\mathsfit}{bold}{\encodingdefault}{\sfdefault}{bx}{n}

\newcommand{\para}[1]{\noindent \textbf{#1}}

\newlength\savedwidth

\author{Yunzhu~Pan$^{*\dagger}$}
\affiliation{%
  \institution{
  University of Electronic Science and Technology of China, Chengdu, China}
  }

\author{Nian~Li$^{*}$}
\author{Chen~Gao$^{\ddagger}$}
\affiliation{%
  \institution{
  Beijing National Research Center for Information Science and Technology, Department of Electronic Engineering, Tsinghua University, Beijing, China}
  }

\author{Jianxin~Chang}
\author{Yanan~Niu}
\author{Yang~Song}
\affiliation{%
  \institution{Beijing Kuaishou Technology Co., Ltd., Beijing, China}
  }

\author{Depeng~Jin}
\author{Yong~Li}
\affiliation{%
  \institution{
  Beijing National Research Center for Information Science and Technology, Department of Electronic Engineering, Tsinghua University, Beijing, China}
  }

\thanks{$*$Both authors contributed equally to this research.\\
$\dagger$Work done when interning at Tsinghua University.\\
$\ddagger$Chen Gao is the corresponding author (chgao96@gmail.com).}

\renewcommand{\authors}{Yunzhu Pan, Nian Li, Chen Gao, Jianxin Chang, Yanan Niu, Yang Song, Depeng Jin, Yong Li}
\renewcommand{\shortauthors}{Yunzhu Pan et al.}

\copyrightyear{2023}
\acmYear{2023}
\setcopyright{acmlicensed}\acmConference[CIKM '23]{Proceedings of the 32nd ACM International Conference on Information and Knowledge Management}{October 21--25, 2023}{Birmingham, United Kingdom}
\acmBooktitle{Proceedings of the 32nd ACM International Conference on Information and Knowledge Management (CIKM '23), October 21--25, 2023, Birmingham, United Kingdom}
\acmPrice{15.00}
\acmDOI{10.1145/3583780.3615482}
\acmISBN{979-8-4007-0124-5/23/10}

\begin{document}

\title{Learning and Optimization of Implicit Negative Feedback for \\ Industrial Short-video Recommender System}

\begin{abstract}
Short-video recommendation is one of the most important recommendation applications in today's industrial information systems.
Compared with other recommendation tasks, the enormous amount of feedback is the most typical characteristic.
Specifically, in short-video recommendation, the easiest-to-collect user feedback is the \textit{skipping} behavior, which leads to two critical challenges for the recommendation model.
First, the skipping behavior reflects implicit user preferences, and thus, it is challenging for interest extraction.
Second, this kind of special feedback involves multiple objectives, such as total watching time and skipping rate, which is also very challenging.
In this paper, we present our industrial solution in Kuaishou\footnote{\url{www.kuaishou.com}}, which serves billion-level users every day.
Specifically, we deploy a feedback-aware encoding module that extracts user preferences, taking the impact of context into consideration. 
We further design a multi-objective prediction module which well distinguishes the relation and differences among different model objectives in the short-video recommendation.
We conduct extensive online A/B tests, along with detailed and careful analysis, which verify the effectiveness of our solution.

\end{abstract}



\maketitle
	
\section{Introduction}\label{sec::intro}

Short-video platforms such as TikTok\footnote{\url{www.tiktok.com}} and Kuaishou have achieved great success, attracting billions of users who spend hours watching videos every day.
Distinct from traditional online recommender systems like news and e-commerce, building an effective short-video recommender system poses a new and challenging task due to its special interaction manner.
In traditional recommender systems, user-item interaction feedback can be categorized into explicit feedback~\cite{mnih2007probabilistic} and implicit feedback~\cite{rendle2009bpr,he2016fast}. 
Explicit feedback mainly refers to the feedback that directly conveys whether the user likes the item or not, such as movie ratings on a scale of 1-5;
implicit feedback refers to binary click data in most applications and only reveals implicit user preferences, not fully describing whether the user likes or dislikes the item. 

Despite the success of various recommendation models that try to teach users implicit or explicit feedback~\cite{wu2022graph,wu2022survey,hu2008collaborative,rendle2021item,rendle2020neural}, the short-video recommendation is required to resolve the different feedback.
Specifically, the short videos are exposed (\textit{i.e.}, recommended) continuously, which leads to a critical challenge that the user feedback is different from existing recommenders, such as the clicks/purchases in e-commerce websites or ratings in movie websites.
That is to say, in such a streaming-form interaction manner of recommender systems, existing methods are not suitable for learning preferences from user feedback.
Moreover, the user-item interaction that is easiest to collect is the \textit{implicit negative feedback}.
Specifically, users can choose to skip over the recommended video, which is always the only feedback we collect for the video since users seldom choose to like or dislike a video, not to mention commenting, sharing, or other behaviors.
Thus, learning from the new form of implicit negative feedback is an important but less-explored problem, which is suffering from two major challenges as follows.

\begin{itemize}[leftmargin=*]
    \item \textbf{Extracting preference signal from feedback.} Users' behaviors will be fuzzier if most behaviors are just implicit skipping behaviors, which leads to unclear user preferences. While preference learning is fundamental to accurate recommendations, learning from plenty of new implicit feedback is challenging.

    \item \textbf{Complex objectives for different feedback are involved.} The implicit skipping-over feedback in short-video recommendations makes the recommender systems involve multiple optimization goals, such as skipping rate minimization, watch time maximization, etc. Thus, the model optimization toward multiple fused objectives is also challenging. 
\end{itemize}

In this paper, we introduce our industrial solution in Kuaishou to address these challenges. 
First, we deploy a feedback-aware sequential encoding module that can introduce multiple feedback, including negative feedback, into the existing context feature extraction. Given the impact of context on the users' decision-making process, the encoding module extracts the reference signal in a context-aware manner.
Second, we deploy a parameter-sharing multi-objective prediction module, which can take context and user/item embeddings as input and well balance the different optimization goals during the prediction.
The shared parameters reveal the relatedness among the goals, while the objective-specific parameters reveal the differences.
In short, our proposed solution can address the above-mentioned critical challenges, providing accurate and personalized recommendations to users on short-video platforms, and is sufficiently evaluated in a real-world deployment.
To summarize, the contribution of this work is as follows.
\begin{itemize}[leftmargin=*]
    \item To our knowledge, we take the early step in approaching the problem of learning and optimization for the new form of implicit negative feedback in industrial short-video recommendations. 
    \item We develop a system that addresses critical and unresolved challenges in preference learning and model optimization. The feedback-aware encoding module well extracts user preference, taking the impact of context into consideration. The multi-objective prediction module well distinguishes the relations and differences among different model objectives in the short-video recommendation.
    \item We deployed our system for billion-level daily users, and extensive A/B tests confirmed its effectiveness.
\end{itemize}

\section{The proposed system}\label{sec::method}

We present the deployed system for the learning and optimization of the new-type negative feedback in Kuaishou.
Our proposed system serves in the ranking phase, producing the final recommendation list.
The system includes three components as follows,
\begin{itemize}[leftmargin=*]
    \item \textbf{Feedback-aware sequential encoder}. We deploy the sequential encoder, which can extract preferences from the mixed user feedback collected from the Kuaishou App in real-time.
    \item \textbf{Context feature embedding layer}. We develop a feature embedding layer that can utilize context information to filter useful prediction signals from sequential embeddings.
    \item \textbf{Multi-feedback prediction layer}. We deploy a multi-task prediction layer with shared parameters to predict multiple types of feedback, especially for the new type of negative feedback.
\end{itemize}

\subsection{Feedback-aware Sequential Encoder}\label{user-encoder}

User feedback, encompassing both positive and negative types, is vital for modeling user preferences. In the real world, negative feedback is often implicit; users tend to skip over disliked content without explicitly expressing their aversion. This subtlety complicates user modeling. To tackle this, we design a feedback-aware encoder that integrates implicit negative feedback with two kinds of positive feedback.

We use video watch time, compared with other users who watched the same video, to define each feedback, including \textbf{E}ngaged \textbf{V}ideo \textbf{V}iewing(\textbf{EVV}), \textbf{F}ocused \textbf{V}ideo \textbf{V}iewing(\textbf{FVV}), and \textbf{G}lance \textbf{V}ideo \textbf{V}iewing(\textbf{GVV}), showed in~\autoref{tab::viewing_type}.
For cold videos without historical user data, user feedback behavior is defined by the same threshold in Table~\ref{tab::viewing_type} based on the total video duration.
\vspace{-3.0mm}
\begin{table}[h]
\caption{Feedback Types: EVV (positive feedback); FVV (stronger positive than EVV); GVV (negative feedback).}
\centering
\vspace{-3.0mm}
\begin{tabular}{cc}
\toprule
\textbf{Feedback Type} & \textbf{Description} \\
\midrule
\textbf{EVV} & Watch time > 50\% of other users. \\ %
\textbf{FVV} & Watch time > 60\% of other users. \\  %
\textbf{GVV} & Watch time < 3 seconds. \\ %
\bottomrule
\end{tabular}
\label{tab::viewing_type}
\end{table}
\vspace{-5.0mm}

\subsubsection{Embedding Layer}
Let $\mathcal{U}$ and $\mathcal{I}$ denote the sets of users and items, respectively. Given a user $u \in \mathcal{U}$, its history behavior sequences is denoted as $\mathbf{S}^u = \{s^u_1, s^u_2, \cdots,  s^u_{|\mathbf{S}^u|}\}$. First, we build an item embedding matrix $\mM \in \mathbb{R}^{|\mathcal{I}| \times d}$, where $d$ denotes the dimension size, and the retrieved corresponding item embedding sequence is as follows,
\begin{equation}
\mE^u = (\ve^u_1, \ve^u_2,  \cdots, \ve^u_{|\mathbf{S}^u|}).
\end{equation}

Since each behavior belongs to multiple types of feedback, we introduce another sequence $\mR^w_u$ to represent each feedback type, EVV, FVV, and GVV, respectively. 
\begin{equation}
\mR^w_u = \{r^w_{u,1}, r^w_{u,2}, \cdots,  r^w_{u,|\mathbf{S}^u|}\}, w \in \{e, f, g\},
\label{eq::feedback}
\end{equation}
where $r^w_{u,i}$ is either 1 or 0, corresponding to True and False.

\subsubsection{Self-Attention Encoder}\label{self-attention}

To capture the relationships between each user behavior in the sequence, we employ a self-attention encoder. Initially, we concatenate the feedback \( \mathbf{R}^w_u \) in Eqn~(\ref{eq::feedback}) with the item embedding \( \mE^u \) and integrate a learnable position embedding \( \mathbf{p} \) for position information. This forms the self-attention encoder input \( \hat{\mE}^{u} \in \mathbb{R}^{L \times (d+3)} \), where \( L \) denotes the sequence length,

\begin{equation}
\hat{\mE}^{u}= [ \mE^{u} ; \mR_{u}^{e} ; \mR_{u}^{f} ; \mR_{u}^{g} ] + \mathbf{p}.
\end{equation}

The core self-attention mechanism is defined as follows,

\begin{equation}
\operatorname{SelfAttention}(\mathbf{Q}, \mathbf{K}, \mathbf{V})=\operatorname{softmax}\left(\frac{\mathbf{Q} \mathbf{K}^T}{\sqrt{d}}\right) \mathbf{V}.
\end{equation}

Each block of the self-attention encoder is constructed as follows,

\begin{equation}
\begin{aligned}
\mathbf{H}^{\prime} & =\operatorname{LayerNorm}\left(\text { SelfAttention }(\mathbf{XW^q}, \mathbf{XW^k}, \mathbf{XW^v})+\mathbf{X}\right), \\
\mathbf{H} & =\operatorname{LayerNorm}\left(\operatorname{FFN}\left(\mathbf{H}^{\prime}\right)+\mathbf{H}^{\prime}\right).
\end{aligned}
\end{equation}

Here, the projection matrices \( \mathbf{W^q}, \mathbf{W^k}, \mathbf{W^v} \in \mathbb{R}^{(d+3) \times (d+3)} \). \( \mathbf{X} \) denotes the output from the preceding layer, with the first layer set to \( \hat{\mE}^{u} \). After processing through the self-attention encoder's final layer, we derive the outputs \( \ve_{h} \), representing the user's sequential behavior history.

\subsection{Context Feature Embedding Layer}\label{context-info}

Besides the user history behavior sequence, context information (e.g., user demographics, item attributes, and platform type) also plays a significant role in learning user preferences. 
For example, young users (\textit{age} attribute of user) may prefer to watch videos about electronics (\textit{category} attributes of video) on the Web platform for a longer time (\textit{platform} type).
Hence, we utilize a feature interaction layer to incorporate the context information, including three phases: feature construction, feature transformation, and deep feature interaction.

\subsubsection{Feature construction.} 
Let $\ve_{u}$, $\ve_{i}$, $\ve_{p}$ be the retrieved embedding of the user, target item, and corresponding platform. We also encode the user's profile information, e.g., age and location, and the video category feature to be the input embedding $\mE_{in} \in \mathbb{R}^{D \times 1}$, where $D$ is the size of the all concatenated features.
\subsubsection{Feature transformation.} 
Context's impact on user behaviors depends on the specific user, item, and platform. 
We design an embedding transform layer to learn the projection relationship between input embedding $\mE_{\text{in}}$ and specific context information $\ve_{u}$, $\ve_{i}$, $\ve_{p}$.
Specifically, we first build a transformation layer as follows,

\begin{equation}
\mathbf{E}_{\text{trans}}^{(u,i,p)} = \mathbf{E_{in}} \mathbf{W}_1 \cdot [\ve_{u}; \ve_{i}; \ve_{p}],
\end{equation}
where the output $\mathbf{E}_{\text{trans}}^{(u,i,p)}$ has the same size of $\mathbf{E_{in}}$. Here superscript $(u,i,p)$ means the embedding transformation depends on a specific user, item, and platform.
That is, $\mathbf{E}_{\text{trans}}^{(u,i,p)}$ contains useful prediction signals, and useless ones are filtered out.

\subsubsection{Feature importance alignment.}
After we obtain $\mathbf{E}_{\text{trans}}^{(u,i,p)}$, which has filtered out useless signals, the importance of different features still varies.
For example, item category is always the most important item feature.
Specifically, the feature importance of learning can be at different granularity.
The finest-grained importance can be dimension-aware, \textit{i.e.}, assigning a weight value to each dimension of $\mathbf{E}_{\text{trans}}^{(u,i,p)}$.
The coarsest-grained importance actually refers to assigning a universal weight to all dimensions, not distinguishing the different feature importance.

To achieve adaptive-grained importance learning, we divide  $\mathbf{E}_{\text{trans}}^{(u,i,p)}$ to $S$ slots (the size of each slot is $D/S$ and $S$ is a hyper-parameter) and assign each slot different important weights ($D/S$ dimensions inner one slot share the same weight). We then deploy a weight learning component as follows,
\begin{equation}
\bm{\alpha}  = \mathrm{sigmoid}( \mathbf{W}_\textbf{slot} [\mathbf{E}_{\text{trans}}^{(u,i,p)}; \ve_{u}; \ve_{i}; \ve_{p}]),
\end{equation}
where $\mathbf{W}_\textbf{slot} \in \mathbb{R}^{S \times 4D}$ is a learnable parameter.

With the learned slot-level feature importance, we combine it with $\mathbf{E}_{\text{trans}}^{(u,i,p)}$ to generate the final feature embeddings as follows:
\begin{equation}\label{eqn::slots}
\begin{aligned}
    \widetilde{\mathbf{E}}^{(u,i,p)} = \bm{\alpha} \otimes \mathbf{E}_{\text{trans}}^{(u,i,p)},
\end{aligned}
\end{equation}
where $\otimes$ denotes the element-wise product.

\begin{figure}[t]
    \centering
    \includegraphics[width=\linewidth]{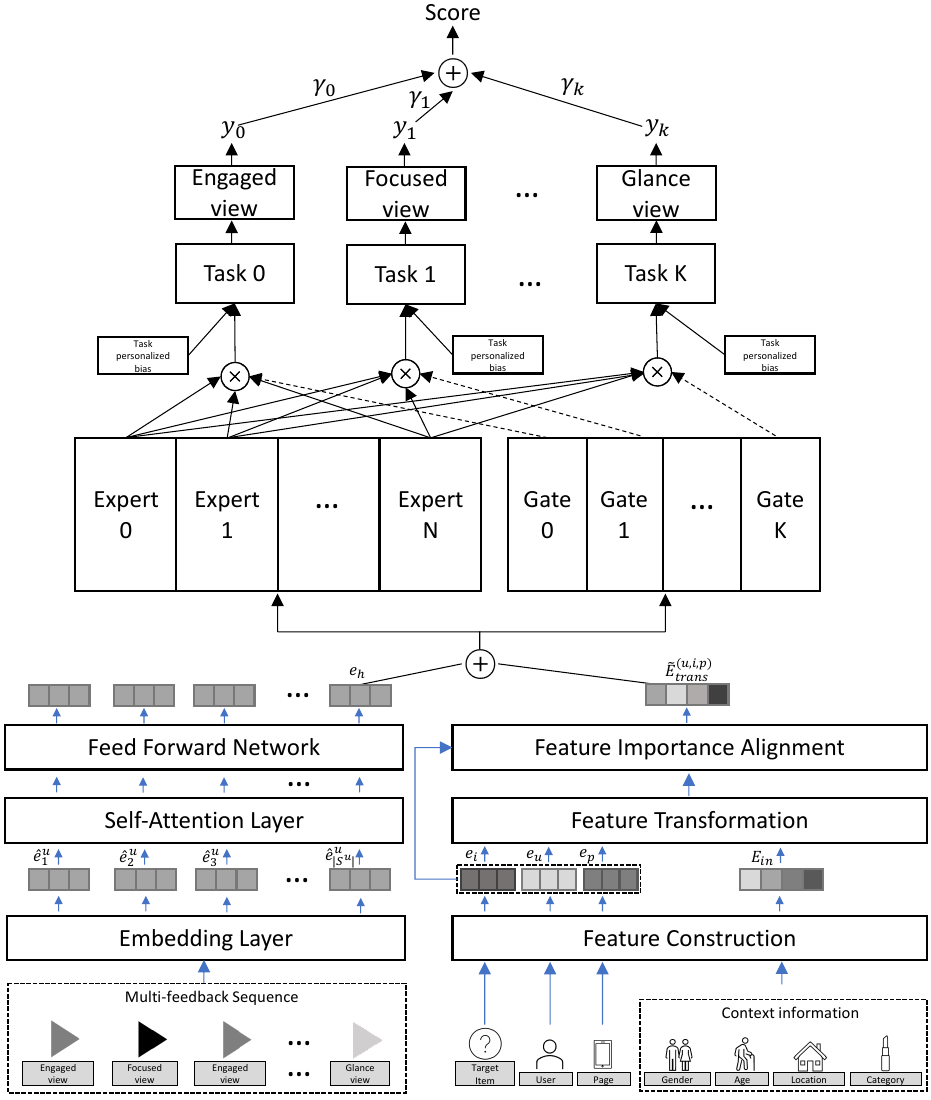}
    \caption{Illustration of our proposed system.}
    \label{fig:model}
    \vspace{-0.6cm}
\end{figure}

\subsection{Multi-feedback Prediction Layer}\label{sec::mmoe}
In real-world recommendation engines, the presence of multiple behaviors poses a multi-objective optimization challenge. For example, a model is optimized to maximize the watch time and minimize the skip ratio at the same time. To address this, we deploy a multi-feedback prediction layer, which takes both user representation $\ve_{h}$, \textit{i.e.}, collaborative filtering signal, and context feature $\widetilde{\mathbf{E}}^{(u,i,p)}$ in 
Eqn~(\ref{eqn::slots}) as input.
That is, it helps learn the behavioral pattern in both historical behaviors and side information.
For clearer presentation, here we denote it as $\mathbf{x} = [\widetilde{\mathbf{E}}^{(u,i,p)}; \ve_{h}]$.

Since multiple behaviors are different but related, we propose a parameter-sharing framework.
We first build $N$ multiple deep neural networks, inspired by~\cite{ma2018modeling}, serving as the shared base network module, denoted as $ f_i = \text{DNN}_i, i = 1,2,\cdots, N, $ 
where $N$ is a hyper-parameter.
Then for $K$ objective-optimization tasks, the corresponding neural network can be a weighted combination of base neural networks. 
In our experimental setup, we assign the value of $K$ as 3, representing the three distinct feedback types: EVV, FVV, and GVV.
Specifically, for the $k$-th task, it is formulated as $ \mathbf{o}_k=\sum_{i=1}^N g^k_i f_i(\mathbf{x}), $
where $g^k_i$ is the $i$-th value of the weight vector $\mathbf{g}^k$, calculated with a gate network as $ \mathbf{g}^k = \text{softmax}(\mathbf{W}^{\textbf{gate}}_k \mathbf{x}), $
where $\mathbf{W}^{\textbf{gate}}_k$ is a trainable parameter.

Based on combining the gate network and base deep neural networks, we obtain the output of each task, $\mathbf{o}_1, \mathbf{o}_2, \cdots, \mathbf{o}_K$.
We then deploy a prediction tower for each task with deep neural networks to obtain the predicted values as $ \mathbf{y}_k = \text{sigmoid}( \mathbf{W}^{\textbf{pred}}_k (\mathbf{o}_k) + b_k), $ 
where $\mathbf{W}^{\textbf{pred}}_k $ and $b_k$ are learnable parameters.

\subsection{Training and Serving}

\subsubsection{Joint training}

The proposed model is optimized in an end-to-end manner, with joint training on $K$ tasks.
Specifically, the training data includes the multi-feedback data denoted as $\mathbb{Y} = \{ \mathbb{Y}_1, \mathbb{Y}_2, \cdots, \mathbb{Y}_K \}$.
For the training data of $k$-th feedback, $\mathbb{Y}_k$, it includes the collected positive samples (observed feedback), $\mathbb{Y}_k^{+}$, and the negative samples randomly selected from unobserved items, $\mathbb{Y}_k^{-}$.
Then the loss function corresponding to the $k$-th feedback can be formulated as follows,
\begin{equation}
    \begin{aligned}
         \mathbb{L}_k = -\mathbb{Y}_k^{+}\log\mathbf{y}_k - (1-\mathbb{Y}_k^{-})\log(1-\mathbf{y}_k),
    \end{aligned}
\end{equation}
where $Logloss$ can also be replaced by other loss functions such as BPR loss (Bayesian Personalized Ranking)~\cite{rendle2009bpr}.

Given the importance of different feedback, in the training process, we adaptively balance tasks with weights to obtain the overall loss function as $ \mathbb{L} = \sum_{k=1}^{K} \lambda_k \mathbb{L}_k $,
where we have $ \sum_{k=1}^{K} \lambda_k = 1 $.

\subsubsection{Fusion strategy for online serving}
Our proposed method serves in the ranking phase of the real-world recommender system, of which the final output is the ranking list exposed to the users.
For the online serving in the real system, we adopt a simple yet effective fusion strategy as 
$ \textbf{score} = \sum_{k=1}^{K}{\gamma_k \mathbf{y}_k}, $
where the selection $\gamma_k$ depends on the real-world requirements; for example, the $\gamma_k$ value for negative feedback should be a negative number.
With the fused score, we generate the $L$-length ranking list from the candidate pool (generated by the previous Recall Phase in the recommendation engine) with the $L$ highest scores.

\section{Online Experiments}\label{sec::result}

In this section, we validate our user negative feedback model via online experiments on Kuaishou, a short-video platform with tens of billions of users.
Despite focusing on a 5\% sample, this still represented hundreds of millions of users. 
Through a 7-day A/B test and real-time data analysis, our approach shows promising results, with detailed findings discussed subsequently.

\subsection{Overall performance of A/B test}

\subsubsection{Experimental Settings}
The details of deployment, baseline, and metrics are presented as follows.

\para{Deployment.}
We conduct an A/B test on Kuaishou's Discover and Featured pages, representing traditional double-column and emerging single-column recommendation scenarios, respectively. These popular pages let us test our model's stability and effectiveness in diverse scenarios while illuminating the role of negative feedback. Hereafter, we'll refer to the Featured and Discover pages as single-column and double-column pages.

\para{Baseline.}
To evaluate the effect of negative feedback, we remove the mixed feedback encoder from the proposed model to serve as the experimental baseline model.
The users are split into two equal groups for an online A/B test: the baseline group receives recommendations from the existing system, and the experimental group receives recommendations from our system that incorporates user negative feedback modeling.

\para{Metrics.}
We monitor real-time user behavior in the A/B test. Metrics represent both user groups' behavior, with the improvement rate calculated as a percentage over the baseline.

\subsubsection{Main results} Our results, detailed in Table~\ref{tab::main_results}, reveal two key insights:
\begin{itemize}[leftmargin=*]
    \item \textbf{By incorporating user negative feedback, our model outperforms the baseline in all measured metrics.} Our model successfully identifies user intent and context-aware preferences, leading to more active users, increasing video viewing time, and enhancing user engagement (evident from more likes and comments), with relative improvements of 0.055\% and 0.160\%, respectively.
    \item \textbf{Integrating implicit user negative feedback, such as Glance Video Viewing, helps reduce explicit negative behaviors.} A notable decrease of 0.049\% users and 0.251\% total reducing times is observed in the use of Kuaishou's "reduce similar videos" function, which is intended to collect explicit negative feedback. Thus, our method effectively handles user feedback, enhancing the user experience.
\end{itemize}

\subsubsection{User Satisfaction Questionnaire Responses}

We further evaluate the efficacy of our method through online satisfaction questionnaires distributed to the experimental and baseline user groups during the A/B test. The questionnaire results in Table~\ref{tab::questionnaire} directly reflect users' preferences.

\textbf{After integrating multiple types of user feedback into the encoding, users' satisfaction with the recommended content increases.} The positive and negative feedback ratios from the questionnaire rise by 1.060\% and fall by 0.089\%, respectively. Simultaneously, we observe a 0.048\% increase in users' likes of the video and a significant reduction in dislikes.
The increase in positive feedback and the decrease in negative feedback indicate that our method infers users' intentions more accurately than the baseline.

\begin{table*}[t!]
    \caption{Main results of A/B performance on user multi-feedback modeling.  Reduction refers to user reports reducing similar recommendations. $\downarrow$ denotes that the lower the value, the more satisfied the user is with the recommended video.}\label{tab::main_results}
    \centering
    \begin{tabular}{ccccccccc}
        \toprule
        Method & \makecell{Active \\ Users} & \makecell{Play \\ Duration} & \makecell{Players \\ Number} & \makecell{Like \\ Users} & \makecell{Like \\ Times} & 
    \makecell{Comment \\ Times} & \makecell{Reduction \\ Users} $\downarrow$ & \makecell{Reduction \\ Times} $\downarrow$ \\
        \midrule
        Improvements rate & +0.055\% & +0.160\% & +0.070\% & +0.046\% & +0.042\% & +0.034\% & -0.049\% & -0.251\%   \\
        \bottomrule
    \end{tabular}
\end{table*}

\begin{table*}[t]
    \caption{The result of the user's response to the user satisfaction questionnaire in the system popup. $\downarrow$ represents negative metrics, in other words, the performance is improved with lower values.}\label{tab::questionnaire}
    \centering
    \begin{tabular}{cccccc}
        \toprule
        Method & \makecell{Questionnaire \\ Pos / Neg feedback} & \makecell{Video \\ Pos / Neg feedback} & \makecell{Video \\ Likes} & \makecell{Questionnaire \\ Dislikes} $\downarrow$  & \makecell{Video \\ Dislikes} $\downarrow$ \\
        \midrule
        Improvements rate & +1.060\% & +0.089\% & +0.048\% & -1.552\% & -0.033\%   \\
        \bottomrule
    \end{tabular}
\end{table*}

\subsection{Performance on different scenarios}

\subsubsection{Comparison between single-column page and double-column page.}

We compare performance between Kuaishou's single-column and double-column pages, which represent short videos and traditional recommendations, respectively. Key metrics analyzed include forward and comment counts, explicit negative behaviors, page visitors, video plays, and reductions in similar recommendations.

As shown in Figure~\ref{tab::comparison}, \textbf{our method outperforms the baseline across all the user-side and video-side metrics on both page types.} The utilization of negative feedback  in our method brings an increased number of page visitors and video plays (growth rates of 0.060\% and 0.079\% for single-column pages; 0.041\% and 0.057\% for double-column pages) and higher user engagement (increases in forwarding and commenting by 0.0194\% and 0.102\% in single-column; 0.776\% and 0.057\% in double-column). This indicates users' positive reception to the recommended content.
Moreover, our method reduces negative user feedback, with fewer users reporting dislikes and reducing similar recommendations. This performance across different scenarios certifies our model's stability and effectiveness.

\begin{table*}[t]
    \centering
    \caption{Performance of our model under single-column page and double-column page scenarios. $\downarrow$ denotes that the lower the value, the better performance our model achieved.}
    \label{tab::comparison}
    \begin{tabular}{ccccccc}
        \toprule
        Page & Forward & Comment & Negative & Visitors & Players & Reduction $\downarrow$ \\
        \midrule
        Single-Column & +0.194\% & +0.102\%	& -0.867\% & +0.060\% & +0.079\% & -0.093\% \\
        Double-Column & +0.776\% & +0.042\% & -2.045\% & +0.041\% & +0.057\% & -2.182\% \\
        \bottomrule
    \end{tabular}
\end{table*}

\begin{figure*}[t]
    \centering
        \subfigure[]{\label{fig::2a}
            \begin{minipage}[h]{0.18\linewidth}
            \includegraphics[width=\linewidth]{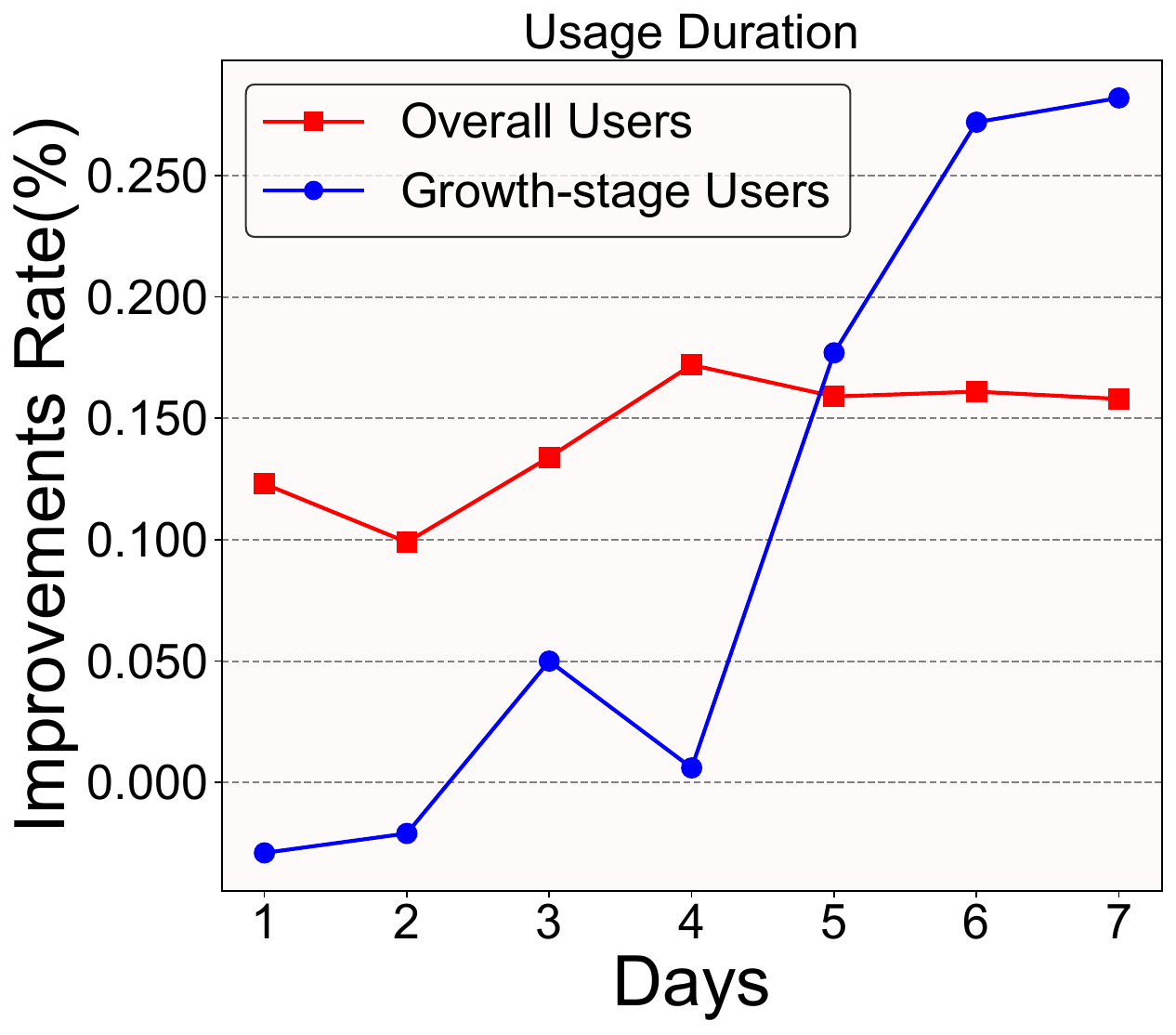}
            \end{minipage}
        }
        \subfigure[]{\label{fig::2b}
            \begin{minipage}[h]{0.18\linewidth}
            \includegraphics[width=\linewidth]{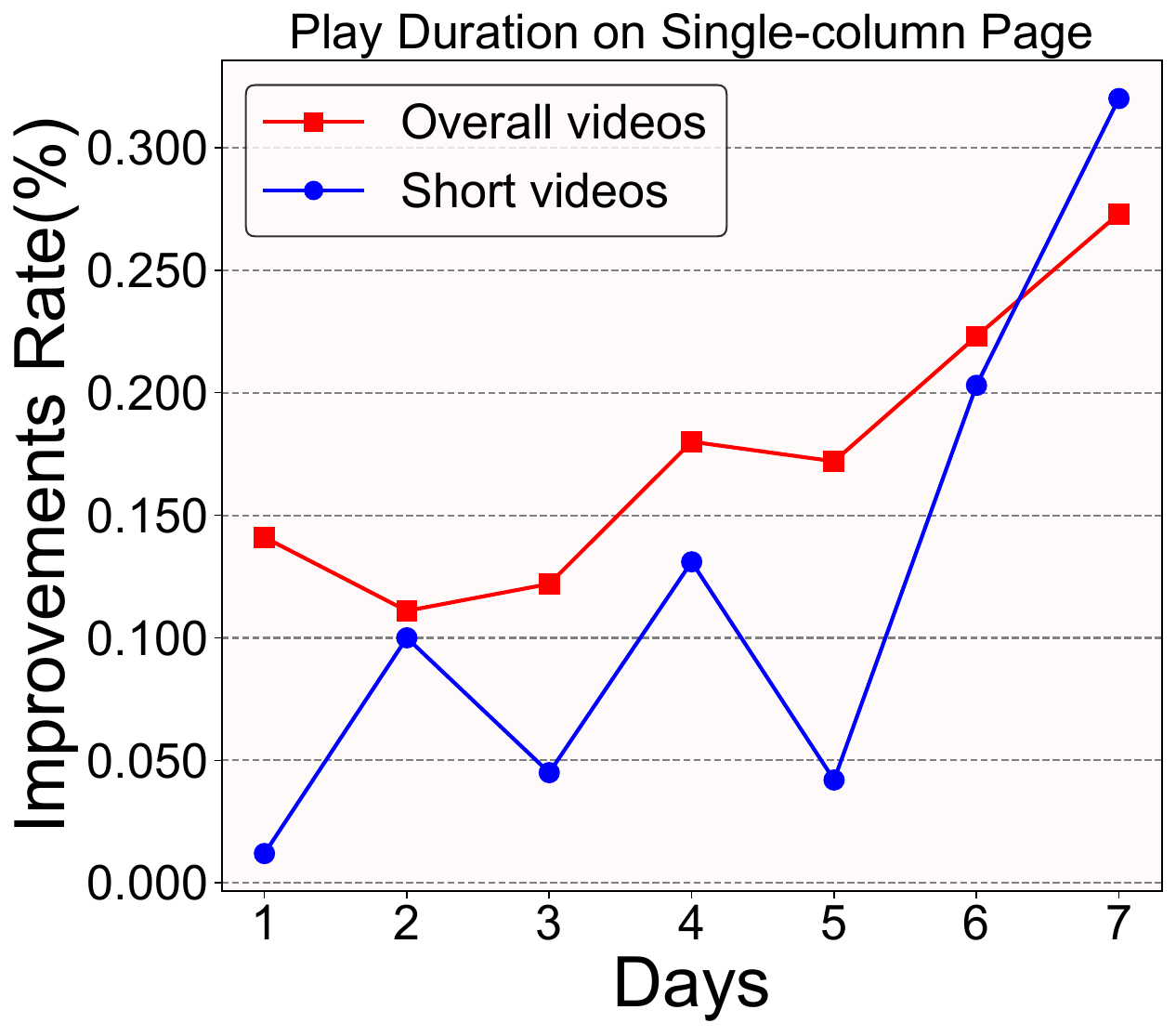}
            \end{minipage}
        }
        \subfigure[]{\label{fig::2c}
            \begin{minipage}[h]{0.18\linewidth}
            \includegraphics[width=\linewidth]{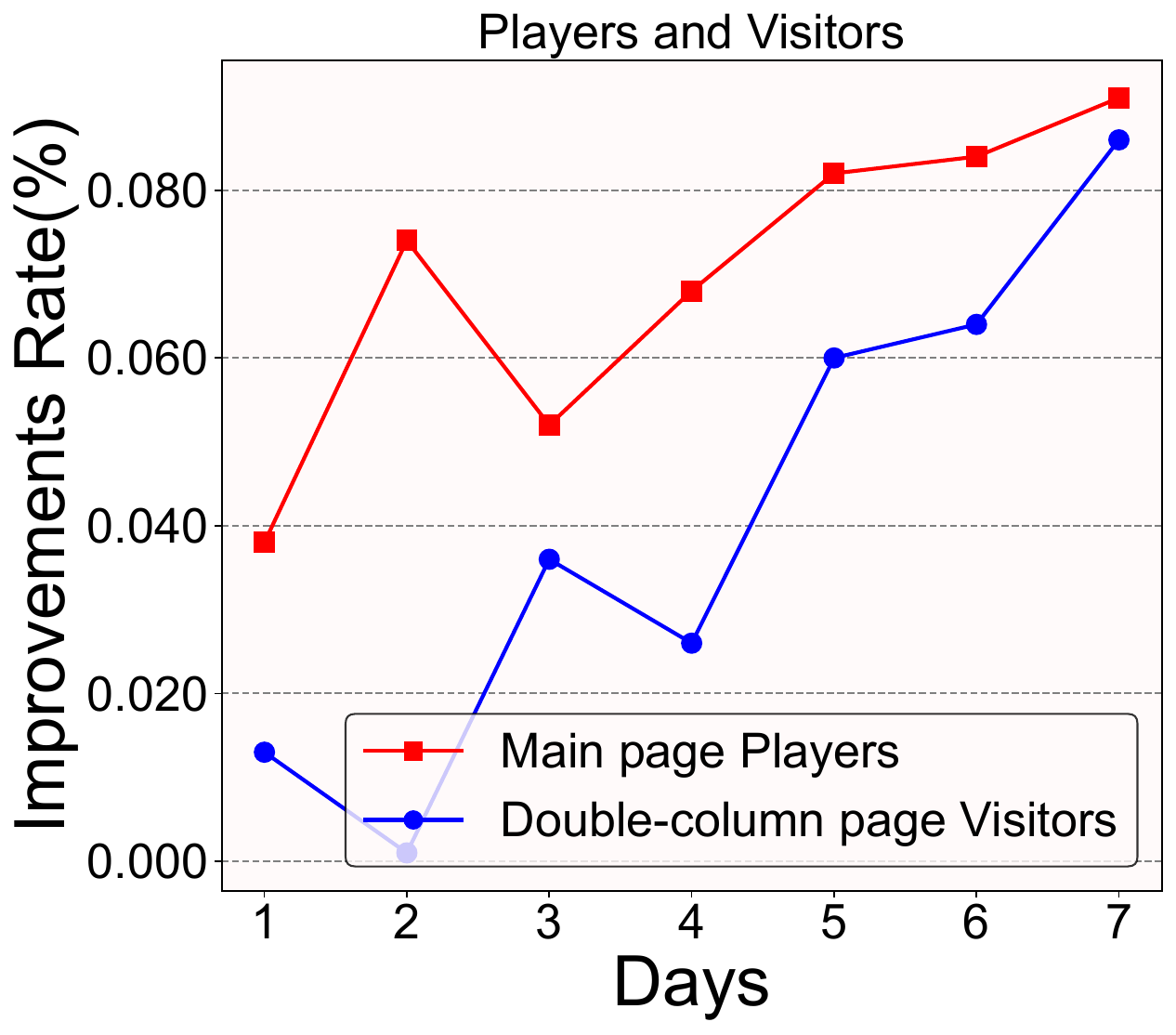}
            \end{minipage}
        }
        \subfigure[]{\label{fig::2d}
            \begin{minipage}[h]{0.18\linewidth}
            \includegraphics[width=\linewidth]{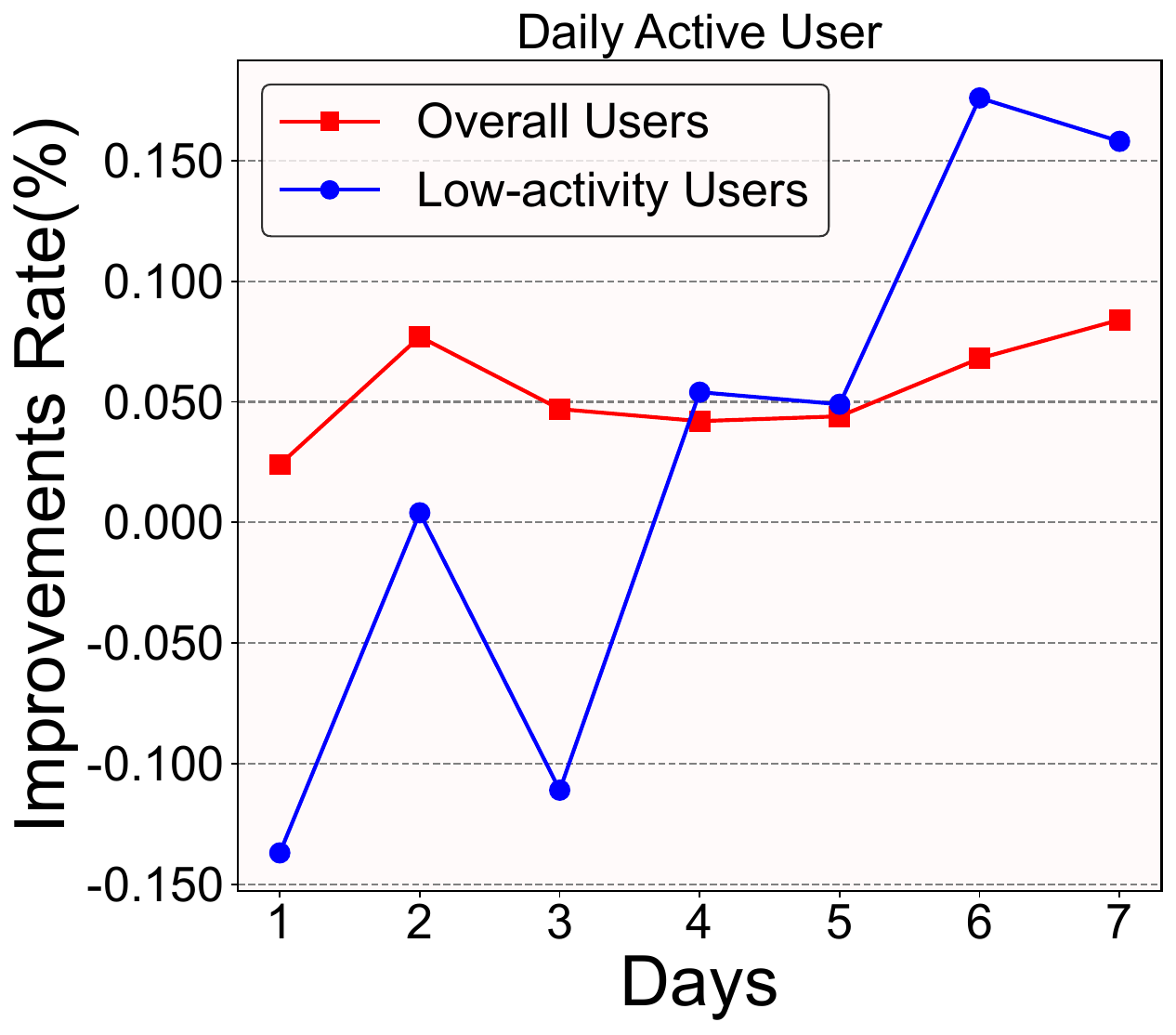}
            \end{minipage}
        }
        \subfigure[]{\label{fig::2e}
            \begin{minipage}[h]{0.18\linewidth}
            \includegraphics[width=\linewidth]{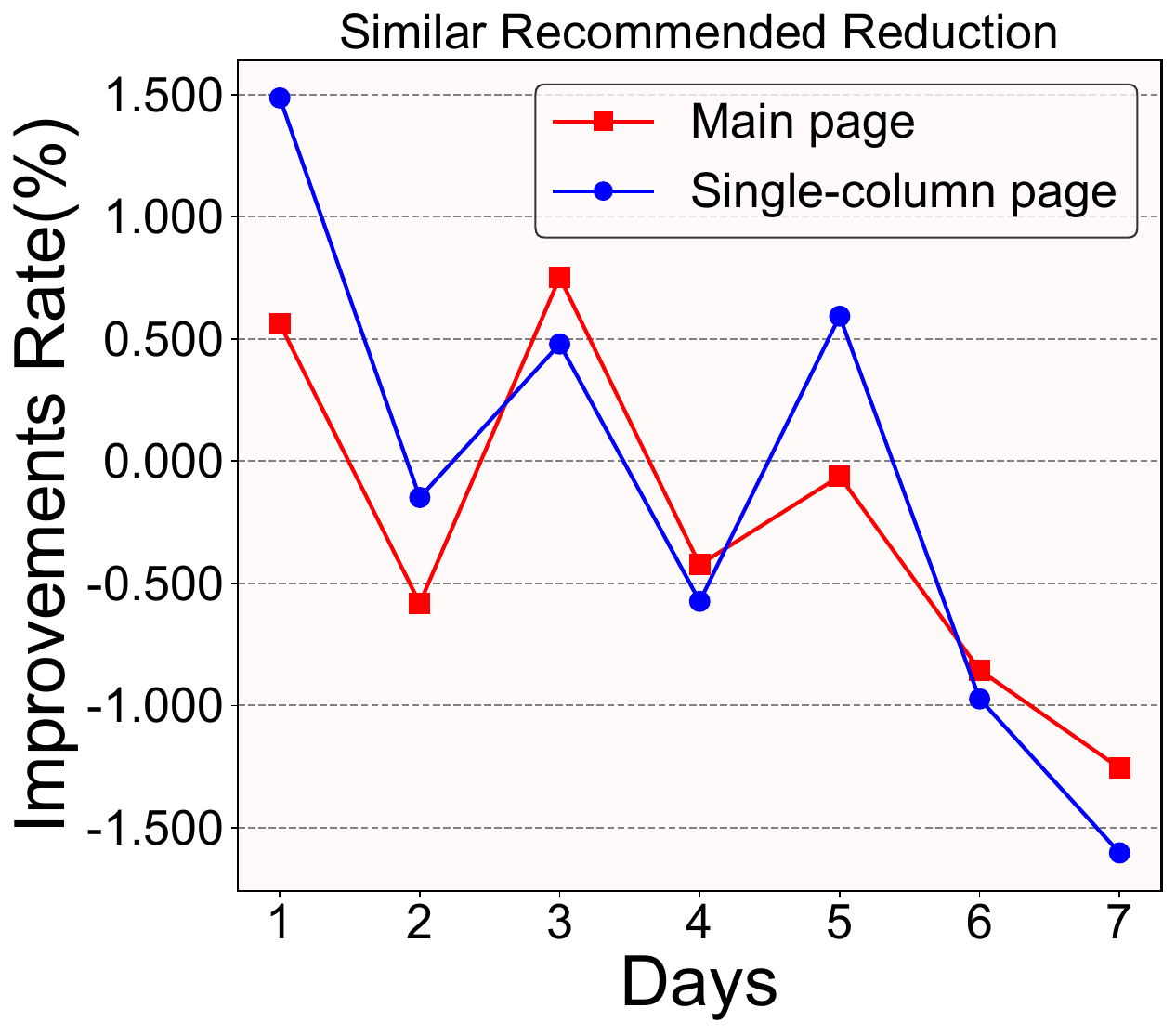}
            \end{minipage}
        }
        \vspace{-0.2cm}
        \caption{The performance improvement trend of our model in a one-week window. (a) The improvement trend of user's usage duration. (b) The improvement trend of video's playing duration on single-column pages. (c) The improvement trend of the main page's players and double-column page's visitors. (d) The improvement trend of the number of daily active users. (e) The improvement trend of the reduction number of similar recommendations, of which the lower number the better performance.}\label{fig::trend}
       \vspace{-0.2cm}
\end{figure*}

\subsubsection{Improvement trend over a week}
we further study the performance improvement in a one-week time window to capture long-term performance improvement. The daily average recommendation performance for our model (bucket-A) and the base model (bucket-B) is presented in Figure~\ref{fig::trend} across key metrics, including app usage duration, video play time and count, active user and page visitor counts, and similar recommendation reductions.
\begin{itemize}[leftmargin=*]
    \item \textbf{The overall app usage duration and the number of active users show stable and smooth improvements over the baseline.} Figure~\ref{fig::2a} shows a steady improvement trend in app usage duration, with an average growth rate of 0.150\%, and relative stability in active user count in Figure~\ref{fig::2d}, with an average improvement rate of around 0.050\%.

    \item \textbf{Our method's improvements increase consistently over time on most metrics.} Growth users' (recently-registered users who are likely to continue using the Kuaishou app) usage time shows an upward trend in Figure~\ref{fig::2a}, with an improvement rate of -0.029\% in the first day to 0.282\% in the last day. 
    Similarly, despite an initial lower count, active user numbers in the low-activity group gradually exceeded the baseline by over 0.15\% as negative feedback modeling proceeds.
    Other metrics show an immediate improvement once our method is deployed. In addition to the increase in positive feedback, explicit negative feedback also decreases, indicating users' growing satisfaction with the recommended videos.
\end{itemize}
The above results show our model's effectiveness and stability in enhancing both short and long-term user engagement.

\vspace{-0.2cm}
\subsubsection{Study on users with different engagement levels}
We investigate whether the observed performance improvement is consistent across users with varying levels of engagement. Users are categorized into three engagement levels: growth stage, maturity stage, and recession stage, with the addition of a low-activity user group.

\begin{figure}[t]
    \centering
    \includegraphics[width=0.47\linewidth]{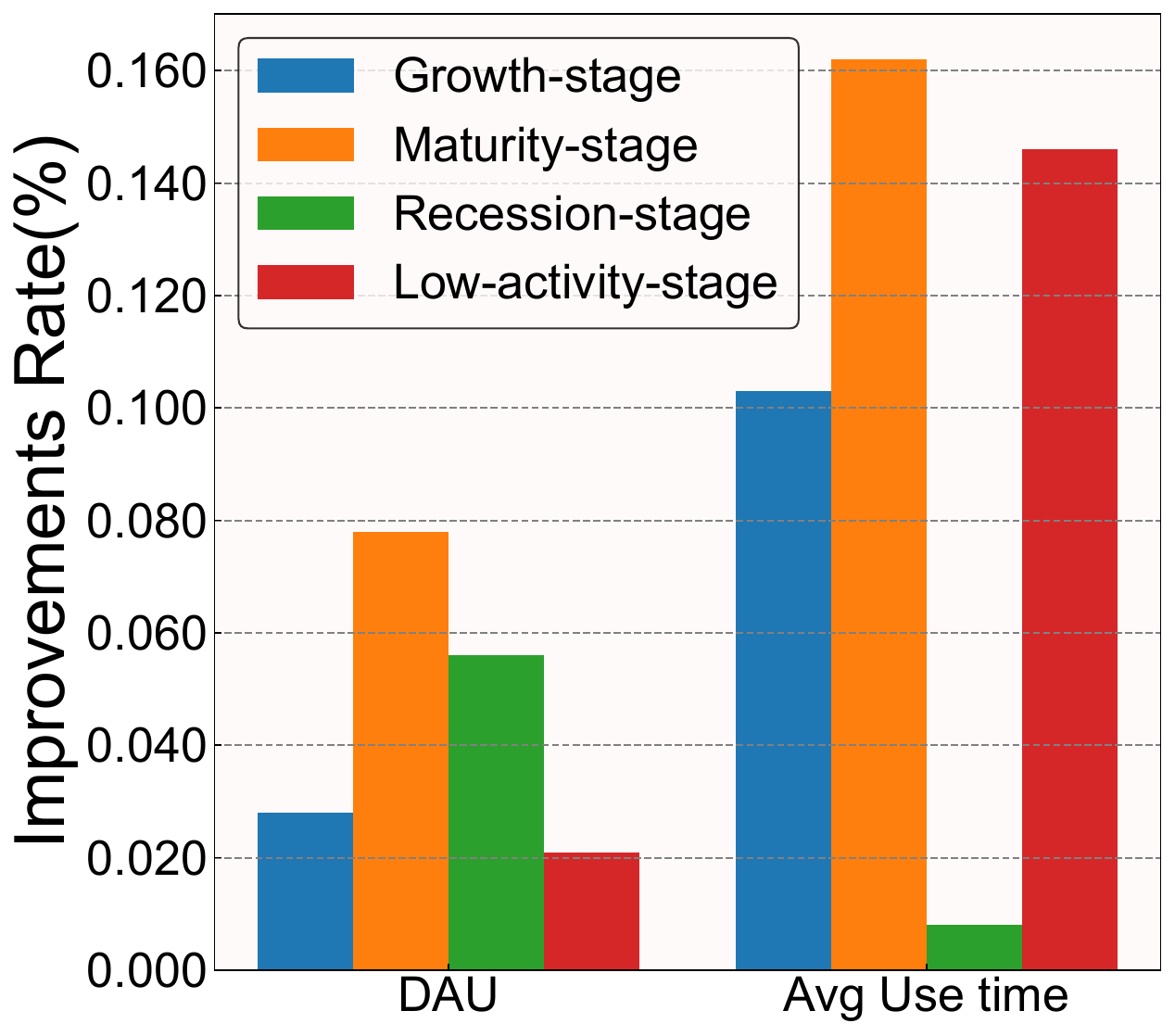}
    \caption{Daily active users (DAU) and user's average using time (Avg Use time) on different user engagement levels.}
    \label{fig:maturity}
\end{figure}

\begin{itemize}[leftmargin=*]
    \item \textbf{Our method outperforms baseline across every group,} showing improvements ranging from 0.008-0.162\% in Figure~\ref{fig:maturity}. This consistent improvement ensures that negative feedback modeling benefits all user groups.
    \item \textbf{The maturity-stage group obtains the most significant improvement.} 
     This is explained by the maturity-stage user group having the most active users and more feedback collected, which further verifies the effectiveness of the feedback modeling.
\end{itemize}

\begin{figure}[t]
    \centering
    \subfigure[Result of low exposure users.]{
        \includegraphics[width=0.42\linewidth]{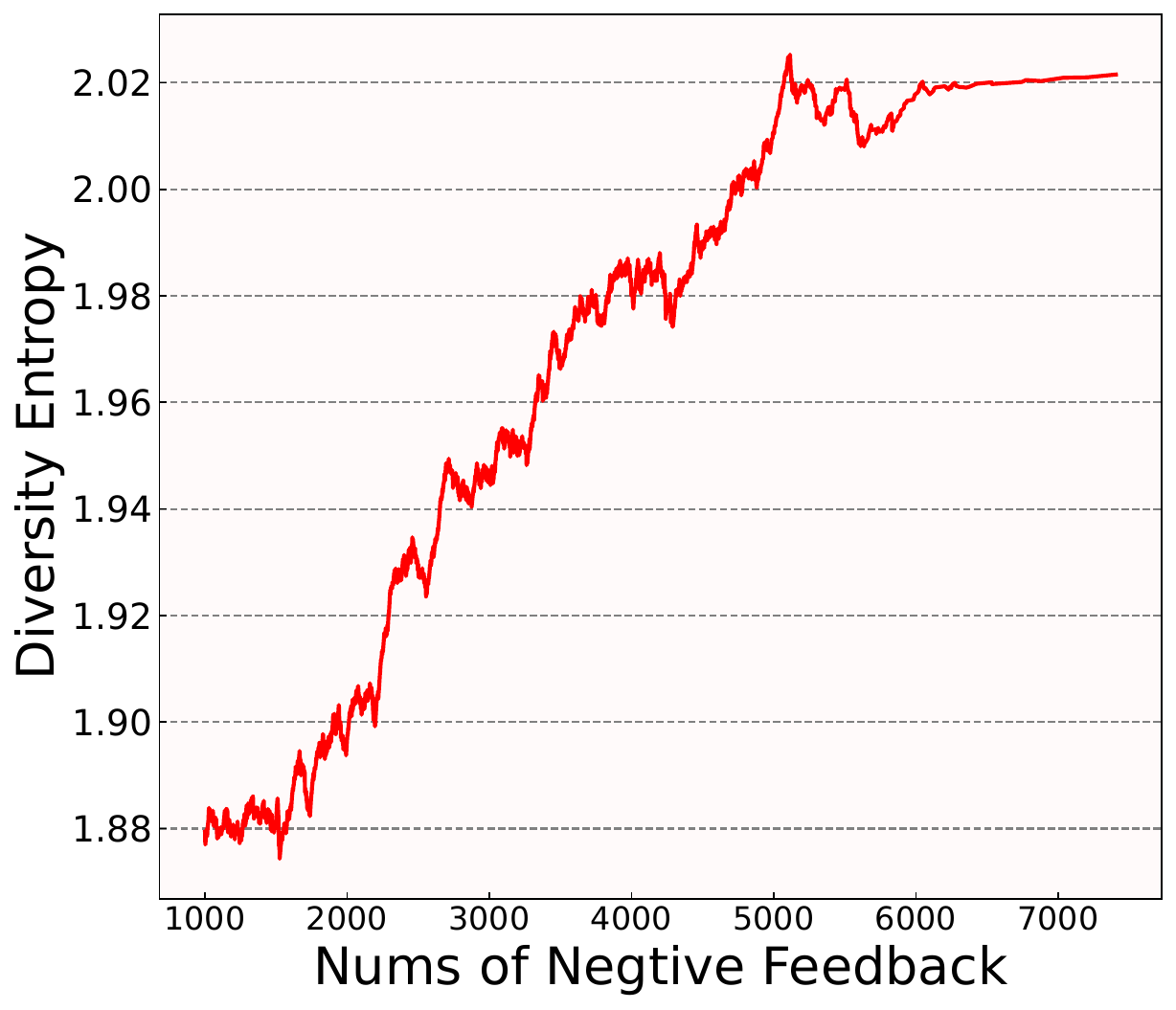}
    }
    \subfigure[Result of high exposure users.]{
	\includegraphics[width=0.43\linewidth]{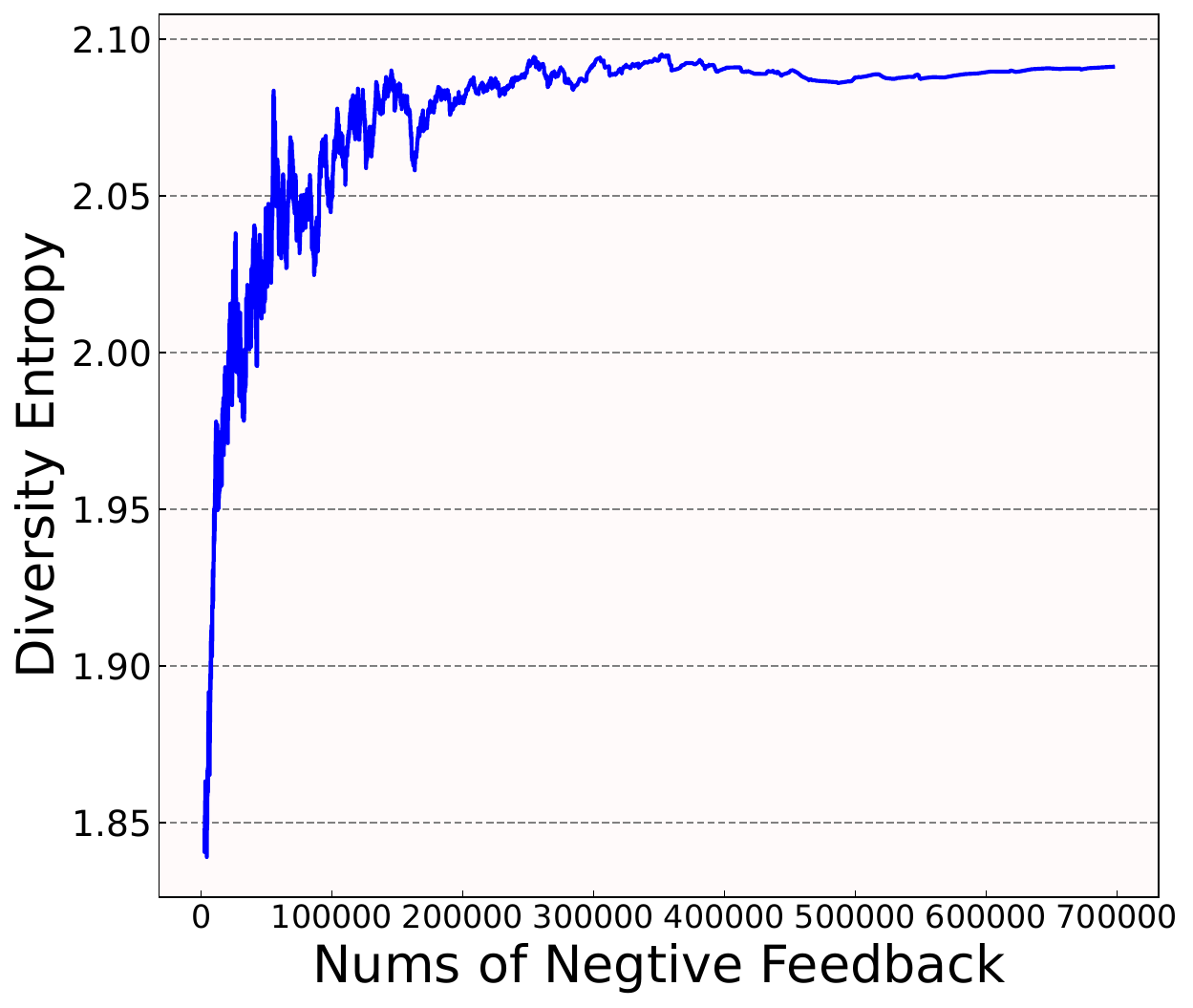}
    }
    \caption{Relationship between negative feedback and diversity entropy in the long-term aspect.}
    \label{fig:longterm}
\end{figure}

\vspace{-2.8mm}
\subsection{Long-term Analysis for Model Deployment}

To further study the long-term influence of our method, we deploy it to the real-world recommender system, Kuaishou, over six months. We record the behaviors of over 600,000 users and investigate the correlation between the diversity of recommendations and users' negative feedback over time. 
We calculate the "diversity entropy~\cite{li2022exploratory}" of each time window containing 100 videos from user history, which evaluates the diversity and randomness of item distribution in a set. For a balanced analysis, users are classified into low-exposure and high-exposure groups. 

As presented in Figure~\ref{fig:longterm}, our model, which effectively incorporates users' negative feedback, enables prompt adjustments for better content diversity over prolonged periods of use.
We can observe that: \textbf{Users who are more willing to express negative feedback will result in a more diverse set of recommended videos by our model.}

This extensive six-month experiment validates the long-term efficacy of our negative feedback modeling approach, demonstrating its potential for steady and significant improvements.

\section{Related Work}\label{sec::relatedwork}

\subsection{Implicit Feedback Learning}

Early recommender systems relied on explicit feedback like ratings~\cite{khan2021deep,koren2009matrix}. However, with the increasing availability of binary interaction data, such as click behavior, implicit feedback learning has become a more common setting for recommendation methods, including early matrix factorization-based methods~\cite{rendle2009bpr,he2016fast,rendle2010factorization,DeepFM,NFM,AFM}, neural network-based ones~\cite{he2017neural,chen2019joint,Cheng2016Wide,xDeepFM,PNN,AutoInt,AutoFIS,AFN,gao2022causal}, advanced graph neural network (GNN) approaches~\cite{he2020lightgcn,wang2019neural,ying2018graph,gao2023survey}, and even to recent automated machine learning ones~\cite{AutoFIS,DARTS}. 
Other works~\cite{ding2020simplify,gong2022positive} proposed a negative sampling strategy to utilize implicit feedback as negative candidates.
However, new and important feedback caused by the streaming-manner recommendation widely exists in the short-video recommendation, such as skipping behavior. While previous works in music streaming ~\cite{meggetto2021skipping, meggetto2023people} studied the types and reasons for skipping behavior, they didn't provide effective solutions  to application problems. However, our system in Kuaishou has well addressed these, which cannot be well handled by existing works since the skipping behavior does not directly reveal user preference signals and involves multiple fused objectives.

\vspace{-2.0mm}

\subsection{Video Recommendation}
Video recommendation, especially for short video apps, is a critical application of recommender models. Existing works ~\cite{deldjoo2016content, huang2016real, mei2011contextual, wei2019mmgcn}mainly focus on how to better exploit the video features into preference learning.
For example, Deldjoo et al.~\cite{deldjoo2016content} proposed to automatically extract visual features of videos and provide users with similar videos based on these video features.
Lee et al.~\cite{lee2017large} utilized raw visual and auditory content to learn a compact representation of videos.
There are some works that approach the problem by studying the optimization watching time~\cite{zheng2022dvr}. Zheng et al.~\cite{zheng2022dvr} proposed an unbiased evaluation metric, watch time gain, to learn unbiased user preferences and alleviate the duration bias in previous evaluation metrics. 
Some other works considered both user and video content attributes to improve recommendation accuracy~\cite{cui2014videos, liu2019user}.
For example, Cui et al.~\cite{cui2014videos} combined social and content attributes and leveraged both social and content information to address the sparsity problem in micro-video recommendation.
Liu et al.~\cite{liu2019user} studied the multi-modal information from both user and short-video sides using the attention mechanism.
However, the challenge of feedback learning, particularly implicit negative feedback for video recommendation, has not been well studied.

\vspace{-2.0mm}
\section{Conclusion and Future Work}\label{sec::conclusion}

In this paper, we present the deployed solution for utilizing implicit negative feedback, which is far easier to collect than other behaviors in the short-video recommendation engine of Kuaishou.
Our solution well combines context-aware feedback learning and multi-objective prediction/optimization, well addressing the two corresponding critical challenges.
The sufficient evaluation through the billion-level users' A/B test demonstrates the effectiveness of the proposed solution.
As for future work, we plan to combine sequential modeling with negative feedback learning, further improving preference learning.

\begin{acks}
This work is partially supported by the National Key Research and Development Program of China under 2022YFB3104702, and the National Natural Science Foundation of China under 62272262, 61972223, U1936217, and U20B2060, and also supported by Kuaishou.
\end{acks}

\balance
\bibliographystyle{ACM-Reference-Format}
\bibliography{bibliography}

\end{document}